\documentclass[doublecol,figures]{epl2} 
\usepackage[dvips]{color}

\begin{document}

\title{Exchange bias effect involved with tunneling magnetoresistance in polycrystalline La$_{0.88}$Sr$_{0.12}$CoO$_3$}

\author{M. Patra \and S. Majumdar \and S. Giri\thanks{E-mail: \email{sspsg2@iacs.res.in}}}

\institute{Department of Solid State Physics, Indian Association for the Cultivation of Science, Jadavpur, Kolkata 700 032, India}

\pacs{85.75.-d}{Magnetoelectronics; spintronics: devices exploiting spin polarized transport or integrated magnetic fields}
\pacs{75.47.De}{Giant magnetoresistance}
\pacs{75.70.Cn}{Magnetic properties of interfaces (multilayers, superlattices, heterostructures)}

\abstract
{We report the exchange bias (EB) effect along with tunneling magnetoresistance (MR) in polycrystalline La$_{0.88}$Sr$_{0.12}$CoO$_3$. Analogous to the shift in the magnetic hysteresis loop along the field ($H$)-axis a shift is clearly observed in the MR-$H$ curve when the sample is cooled in a static magnetic field. Training effect (TE) is a significant manifestation of EB effect which describes the decrease of EB effect when sample is successively field-cycled at a particular temperature. We observe TE in the shift of the MR-$H$ curve which could be interpreted by the spin configurational relaxation model. A strong field-cooled (FC) effect in the temperature as well as time dependence of resistivity is observed below spin freezing temperature. The unusual MR results measured in FC mode are interpreted in terms of intragranular interface effect between short range ferromagnetic clusters and spin-glass matrix giving rise to the EB effect. EB effect in MR has been observed in bilayer or multilayer films which has not yet seen in a polycrystalline compound. EB effect involved with tunneling MR and semiconducting transport property  attributed to the intragranular intrinsic nanostructure is promising for the spintronic applications.}

\maketitle

\section{Introduction}
The exchange bias (EB) effect is a manifestation of the unidirectional anisotropy which is occurred in a system having heterostructure composed of ferromagnetic (FM) and antiferromagnetic substances when it is cooled through the N\'{e}el temperature in a static magnetic field \cite{meik,nogues1,nogues2}. A semiconducting system exhibiting EB effect is fascinating because EB coupling provides an additional degree of freedom for controlling   the conduction process which has a significant impact in spintronic applications  \cite{zutic}. Recently, investigations on the spin polarized transport, tunneling magnetoresistance (TMR), and exchange coupling in the semiconducting FM bilayer or multilayer films have played a central role in spintronics \cite{zutic1,Sanvito}.  

	The doped cobaltites La$_{1-x}$Sr$_x$CoO$_3$ with perovskite structure are the model systems for the investigation on complex magnetoelectronic phase separation \cite{wu1}. At a low doping the semiconducting transport properties were reported where FM metallic clusters embedded in a non-FM insulating matrix have been confirmed by neutron and NMR results  \cite{wu2,kuhns,hoch}. With increasing $x$ the fraction of the FM metallic cluster increases and it coalesces together for $x >$ 0.18 leading to the long range FM ordering and metallic conductivity. Wu {\it et al}. recently reported an interesting scenario of semiconducting transport properties exhibiting glassy magnetic behavior  in the single crystals of La$_{1-x}$Sr$_x$CoO$_3$ for $x \leq$ 0.18 \cite{wu3}. They suggested that the unusual glassy transport phenomena were attributed to the magnetoelectronic phase separation into nanoscale FM clusters embedded in a non-FM matrix. Recently, significant EB effect through the shift in magnetic hysteresis (MH) loop has been reported in  polycrystalline as well as single crystalline compounds,  La$_{1-x}$Sr$_x$CoO$_3$ suggesting the spontaneous phase separation between FM and spin-glass (SG) like states \cite{tang1,tang2,tang3}. 

\begin{figure}[t]
\vskip 0.4 cm
\centering
\includegraphics[width = 7 cm]{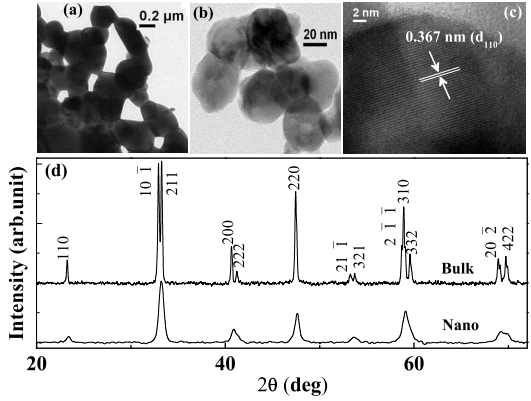}
\caption {TEM image of bulk (a) and nanoparticle (b). (c) HRTEM image of the nanoparticle exhibiting a single (110) plane. (d) Powder x-ray diffraction patterns for the bulk and nanoparticle.}
\label{figure 1}
\end{figure}

Recent reports demonstrate that EB effect can also be evidenced through the shift in  magnetoresistance-field (MR-$H$) curve for bilayer or multilayer films  \cite{dai,niko,kerr,bea}. To the best of our knowledge, signature of EB effect in the MR-$H$ curve is not reported in a compound. In this letter, we report considerable EB effect through the systematic shift, training effect, strong field-cooled (FC) effect in the temperature and time dependences of resistivity ($\rho$) below spin freezing temperature in polycrystalline La$_{0.88}$Sr$_{0.12}$CoO$_3$. Large MR at low temperature is interpreted in terms of intragranular tunneling mechanism across the magnetic tunnel barrier.  Current research has been focused in developing the advanced materials for the spintronic applications in the artificial heterostructure derived from FM semiconductors \cite{maek}. Here, we demonstrate that intrinsic heterostructure attributed to the spontaneous phase separation in the grain interior may provide a useful platform in searching new materials for the spintronic applications.

\begin{figure}[t]
\vskip 0.4 cm
\centering
\includegraphics[width = 7 cm]{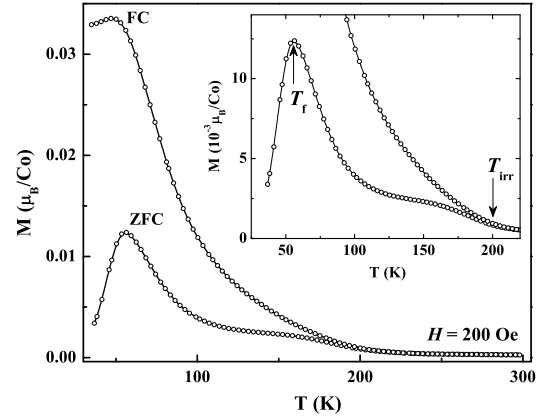}
\caption {Temperature dependence of ZFC and FC magnetizations for the bulk. Inset highlights ZFC magnetization.}
\label{figure 2}
\end{figure}

\section{Experimental}
Polycrystalline compound with composition La$_{0.88}$Sr$_{0.12}$CoO$_3$ was prepared by the chemical route which was described in our previous report \cite{de1}. As-synthesized sample was heated at 873 and 1273 K where the average grain sizes are found to be $\sim$ 35 and $\sim$ 240 nm, respectively which were confirmed by  
Transmission Electron Microscopy (TEM) using a microscope (ZEOL JEM-2010). For simplicity we address the sample with larger grain size as bulk while the sample with smaller grain size is defined as nanoparticle. TEM images are shown in figs. 1(a) and 1(b) for the bulk and nanoparticle, respectively. A high resolution TEM (HRTEM) image of the nanoparticle is further shown in fig. 1(c) where a single lattice plane is noticed with spacing $\sim$ 0.367 nm which matches with the spacing of the (110) plane with rhombohedral structure ($R\overline{3}c$) having lattice parameters, $a$ = 0.538 nm and $\alpha$ = 60.4$^0$ obtained from the x-ray powder diffraction. Experimental results for the bulk is mainly presented here while the results for the nanoparticle is compared with the bulk in fig. 3. Single phase of the rhombohedral structure was confirmed by a powder x-ray diffractometer (Seifert XRD 3000P) using CuK$_{\alpha}$ radiation. Powder x-ray diffraction patterns of the bulk and nanoparticle are shown in fig. 1(d) where the diffraction peaks could be indexed in the rhombohedral structure ($R\overline{3}c$) with lattice parameters, $a$ = 0.539 nm and $\alpha$ = 60.7$^0$ for the bulk which are in accordance with the previous reports \cite{wu1,krien}. Measurements of the negative MR defined as ($\rho_H$ - $\rho_0$)/$\rho_0$ were carried out using a superconducting magnet system (Cryogenic Ltd., UK), where $\rho_H$ and $\rho_0$ are the resistivities in static and zero magnetic field. dc magnetization was measured using a commercial superconducting quantum interference device, SQUID magnetometer (MPMS, XL). 

\section{Results and discussions}
Zero-field cooled (ZFC) and field-cooled (FC) magnetizations as a function of temperature ($T$) measured at 200 Oe are shown in fig. 2 for the bulk. A peak in the ZFC magnetization is observed around $\sim$ 55 K ($T_{\rm f}$) with a large FC effect below $\sim$ 197 K ($T_{\rm c}$) which is highlighted in the inset of the figure. The magnetization results are in excellent agreement with the phase diagram proposed by Wu {\it et al}. where $T_{\rm c}$ is the onset of Curie temperature involved with the short range FM clusters and $T_{\rm f}$ is the spin freezing temperature behaving like a SG transition temperature \cite{wu1}. 

\begin{figure}[t]
\vskip 0.4 cm
\centering
\includegraphics[width = 7 cm]{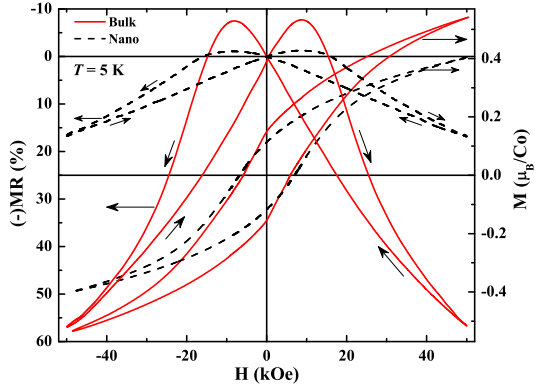}
\caption {(Color online) MH loops and MR-$H$ curves at 5 K for bulk (solid line) and nanoparticle (broken lines).}
\label{figure 3}
\end{figure}

A symmetric MH loop (solid line) measured at 5 K is shown in fig. 3 for the bulk where magnetization does not show any saturating tendency at 50 kOe which is consistent with the cluster-glass state proposed for La$_{0.88}$Sr$_{0.12}$CoO$_3$ \cite{tang2}. MR-$H$ curve (solid line) at 5 K having large MR ($\sim$ 57 \%) at 50 kOe is also displayed in fig. 3 for the bulk. MR (broken line) was also measured for the nanoparticle where magnitude of MR is decreased  drastically than that of the bulk counterpart. Since grain boundary region is considerably enhanced by decreasing the grain size than its bulk counterpart, the large decrease in the magnitude of MR for the nanoparticle reveals that the grain boundary effect does not contribute any significant role where grain interior mechanism leads to the large MR. Intragranular giant magnetoresistance (GMR) attributed to the spontaneous phase separation has been proposed by Wu {\it et al}.  where tunneling between FM metallic clusters across the non-ferromagnetic semiconducting matrix was pointed out for the  interpretation of GMR effect in La$_{0.85}$Sr$_{0.15}$CoO$_3$ \cite{wu2}.
 The results are analogous to the negative intergranular MR previously observed in 
artificial structures composed of nanoscale FM particles embedded in an insulating or metallic nonmagnetic matrix where orientation of the magnetization axes, the density, and the size of the FM entities are crucial to interpret GMR effect \cite{xia,bar,sankar}. In the present observation the smaller magnitude of MR at 5 K and 50 kOe for the nanoparticle than its bulk counterpart is suggested due to the intragranular mechanism involved with the size and the density of FM metallic clusters in the nonferromagnetic semiconducting matrix for the nanoparticle which are different from the bulk analogue.  
MR-$H$ curve attributed to the tunneling mechanism exhibits a peak in MR at zero magnetization \cite{maek}. For example, Co$_{16}$Cu$_{84}$ alloy exhibiting SG-like behaviour at low temperature shows a peak in the MR-$H$ curve at zero magnetization of the magnetic hysteresis loop where magnetic field at the peak in the MR-$H$ curve provides the coercivity \cite{xia}. Here, a peak is observed at $H_{\rm C}^\prime$ (8.8 kOe) which is much larger than the coercivity ($H_{\rm C}$ = 5.8 kOe) in the MH loop. The results are similar to that observed in the double perovskite, Sr$_2$FeMoO$_6$ \cite{sarma}. MH loop and MR-$H$ curve for the nanoparticle are illustrated in fig. 3 by the broken lines. $H_{\rm C}$ is increased to 6.8 kOe while $H_{\rm C}^\prime$ does not change significantly. Since the anisotropy of SG spins is much stronger than FM spins, the coercivity of SG compound is typically much higher than the  FM compound. Thus the system composed of FM and SG components should have smaller average coercivity than the individual SG component where $H_{\rm C}$ obtained from the MH loop provides the average coercivity. In the present observation TMR is attributed to the intragranular effect where tunneling between FM clusters takes place through the SG matrix. In such a case tunneling barrier is set by the anisotropy of the SG spins where $H_{\rm C}^\prime$ represents the coercivity of the SG component. 
We note that the value of $H_{\rm C}^\prime$ does not change with the average particle size. The results further indicate that the anisotropy of the SG component is remained unaltered, despite the grain interior nanostructure is modified due to the change in particle size which is reflected in the substantial change in the MR-$H$ curve.  

\begin{figure}[t]
\vskip 0.4 cm
\centering
\includegraphics[width = 7 cm]{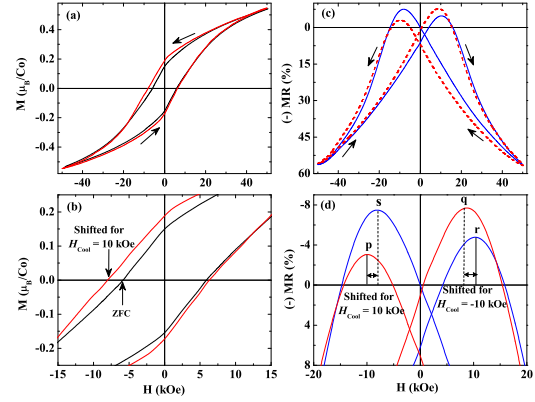}
\caption{(Color online) MH loops (a) for $H_{\rm cool}$ = 0 and 10 kOe (broken curves) and MR-$H$ curves (c) for $H_{\rm cool}$ = 10 (broken curves) and -10 kOe for the bulk. Central part of the MH loops and MR-$H$ curves are highlighted in (b) and (d), respectively.}
\label{figure 4}
\end{figure}

When sample was cooled down to 5 K from 250 K with a cooling field, $H_{\rm cool}$ = 10 kOe, MH loop is shifted along the negative $H$-axis. The shift is absent while cooling the sample in ZFC mode which is shown in fig. 4(a). The central part of the loop is highlighted in fig. 4(b). If $H_{\rm c1}$ and $H_{\rm c2}$ are the negative and positive coercivities of the shifted loop, the magnitude of EB field ($H_{\rm E}$) is defined as $H_{\rm E}$ = $\left|H_{\rm c1} -  H_{\rm c2}\right|$/2 $\approx$ 650 Oe which is consistent with the previous report \cite{tang2}. Interestingly, a large shift in the MR-$H$ curve is also observed at 5 K for cooling the sample in FC mode which is absent in ZFC mode. We further notice that the shift is negative for $H_{\rm cool}$ = 10 kOe while it is positive for $H_{\rm cool}$ = -10 kOe. The shifted MR-$H$ curves in between $\pm$ 50 kOe are illustrated in fig. 4(c) for $H_{\rm cool}$ = $\pm$ 10 kOe which are further highlighted in fig. 4(d) around the origin.  The peak in the MR-$H$ curve measured from 50 kOe to -50 kOe is shifted along negative $H$-axis while the curve from -50 kOe to 50 kOe is remained almost unchanged [see the broken curve in fig. 4(d)] for $H_{\rm cool}$ = 10 kOe. Exchange bias field ($H_{\rm E}^{\rm MR}$) obtained from the shifted MR-$H$ curve is estimated as $H_{\rm E}^{\rm MR}$ = $\left|H_{\rm p} -  H_{\rm q}\right|$/2 $\approx$ 940 Oe where $H_{\rm p}$ and $H_{\rm q}$ correspond to the field of the peak positions at $p$ and $q$, respectively. $H_{\rm E}^{\rm MR}$ is noted much larger than the value from MH loop, despite the values of $H_{\rm E}/H_{\rm C}$ ($\approx$ 11.2 \%) and $H_{\rm E}^{\rm MR}/H_{\rm C}^\prime$ ($\approx$ 10.7 \%) are nearly same. The qualitative interpretation of larger value of $H_{\rm E}^{\rm MR}$ than that of  $H_{\rm E}$ is discussed at the end of the text. We further note that  peak height in the MR-$H$ curve at $p$ is decreased considerably compared to the height at $q$ where peak position and height at $q$ are almost same compared to the symmetric MR-$H$ curve obtained  in the ZFC mode. On the other hand, the peak height at $r$ is decreased considerably than the height at $s$ for $H_{\rm cool}$ = -10 kOe. The results clearly exhibit the spin valve-like character in the MR-$H$ curve analogous to that observed in films exhibiting EB effect \cite{niko,dai,bea,kerr}.


\begin{figure}[t]
\vskip 0.4 cm
\centering
\includegraphics[width = 7 cm]{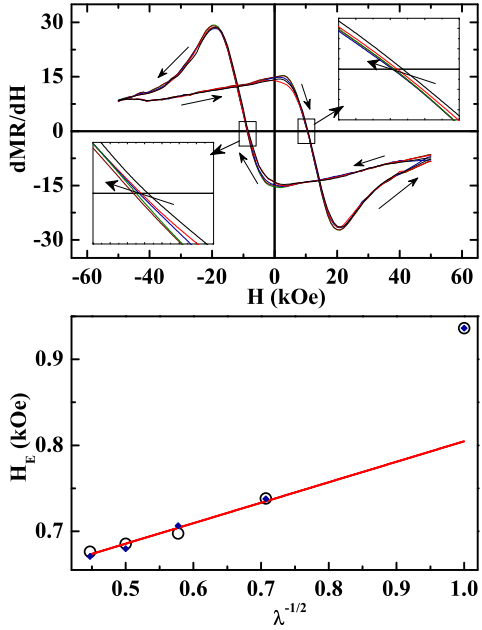}
\caption {(a) Training effect is shown in $d$MR/$dH$ with $H$ plots up to $\lambda$ = 5 for the bulk. (b) Plot of $H_{\rm E}$ with $\lambda^{-1/2}$ where solid straight line exhibits the fit by a power law and filled circles show the calculated $H_{\rm E}$ at different $\lambda$ using Eq. (1). Insets of (a) highlight the curves passing through $d$MR/$dH$ = 0 in negative and positive $H$-axis. Arrows indicate the increase of $\lambda$.}
\label{figure 5}
\end{figure}

Training effect (TE) is one of the significant results of EB effect which describes the decrease of EB effect when the sample is successively field-cycled at a particular temperature. TE is typically observed in the MH loop \cite{nogues1,nogues2} which is recently reported in the MR-$H$ curve only in a very few spin valve systems exhibiting EB effect \cite{ven,brems}. In fig. 5(a) typical signature of TE is illustrated in 
$d$MR/$dH$ with $H$ plots at 5 K up to 5 successive cycles ($\lambda$). Curves passing through $d$MR/$dH$ = 0 in the positive and negative $H$-axes are highlighted in top and bottom inset, respectively which correspond to the field at the peak position in the shifted MR-$H$ curves. 
A large decrease of $H_{\rm E}^{\rm MR}$ $\sim$ 22 \% is observed in between first and second cycles. The decrease of $H_{\rm E}^{\rm MR}$ is fitted satisfactorily with the empirical relation, 
$H_{\rm E}^{\rm MR}(\lambda) - H_{\rm E}^{\rm MR}(\lambda = \infty) \propto \frac{1}{\sqrt{\lambda}}$. 
The solid straight line in fig. 5(b) exhibits the best fit of $H_{\rm E}^{\rm MR}$ against $\lambda^{-1/2}$ ($\lambda \geq$ 2) with $H_{\rm E}^{\rm MR}(\lambda = \infty)  \approx$ 566.5 Oe. The above empirical relation does not fit the sharp decrease between first and second cycles in accordance with the reported results in MH \cite{nogues2,tang1} as well as in MR-$H$ \cite{ven}. Binek proposed a recursive formula in the framework of spin configurational relaxation to understand the training effect which describes the ($\lambda$+1)th loop shift with the $\lambda$th one as \cite{binek}
\begin{equation}
H_{\rm E}^{\rm MR} (\lambda + 1) - H_{\rm E}^{\rm MR} (\lambda) = -\gamma [H_{\rm E}^{\rm MR} (\lambda) - H_{\rm E}^{\rm MR\prime}(\lambda = \infty)]^3
\end{equation} 
where $\gamma$ is a sample dependent constant. Using $\gamma$ = 2.5 $\times$ 10$^{-6}$ Oe$^{-2}$ and $H_{\rm E}^{\infty\prime}$ = 503.5 Oe the whole set of data (filled circles) could be generated which matches satisfactorily with the experimental data [fig. 5(b)]. In case of spin configuration relaxation model \cite{binek} the expression was developed in the frame work of nonequilibrium thermodynamics where TE for the FM/AFM heterostructure was correlated with the relaxation of the FM spins exchange coupled with the AFM spins toward equilibrium. Similar effect might also  be proposed here where the relaxation of the FM spins exchanged coupled with the SG spins at the interface toward equilibrium leads to the TE in MR-$H$ curve.

\begin{figure}[t]
\vskip 0.4 cm
\centering
\includegraphics[width = 7 cm]{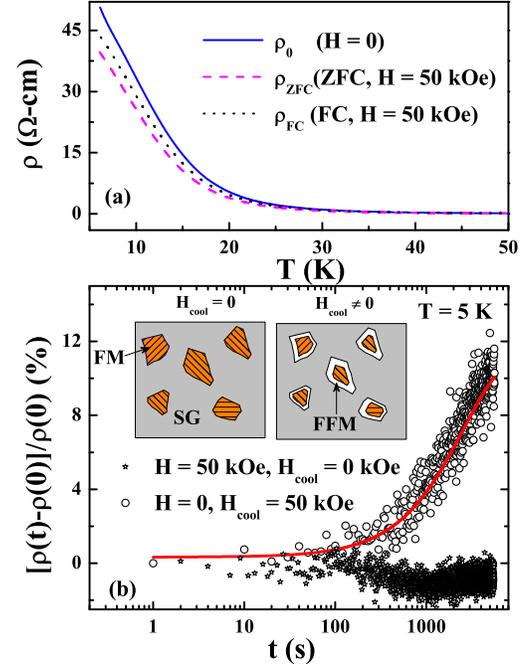}
\caption {(Color online) (a) Temperature variation of resistivity in zero field, $\rho_0$  (solid line) and with field (50 kOe) in ZFC, $\rho_{\rm ZFC}$ (dashed line) and FC, $\rho_{\rm FC}$ (dotted line) modes for the bulk. (b) Time dependence of [$\rho$($t$) - $\rho$(0)]/$\rho$(0) at 5 K measured with $H$ = 0 and 50 kOe in FC and ZFC modes, respectively for the bulk. Cartoon of the phase separation scenario within a grain in ZFC ($H_{\rm cool}$ = 0) (left inset) and FC ($H_{\rm cool} \neq$ 0) (right inset) modes. FM, FFM, and SG represent the ferromagnetic, frozen FM and spin-glass regions.}
\label{figure 6}
\end{figure}

Semiconducting temperature dependence of $\rho$ measured during warming mode in zero field ($\rho_0$) and in field (50 kOe) under ZFC ($\rho_{\rm ZFC}$) and FC ($\rho_{\rm FC}$) modes are shown in fig. 6(a). $\rho_{\rm ZFC}$ deviates from $\rho_0$ below $\sim$ 40 K which is much below $T_{\rm f}$ at 55 K. Furthermore, a considerable deviation between $\rho_{\rm ZFC}$ and $\rho_{\rm FC}$ is observed where the magnitude of $\rho_{\rm FC}$ is significantly higher than $\rho_{\rm ZFC}$ below $\sim$ 40 K. A strong FC effect in the time  dependence of $\rho$ is observed under following experimental protocol. Sample was cooled down to 5 K from 250 K with $H_{\rm cool}$ = 50 kOe and then $\rho$ was measured with time ($t$) after removal of magnetic field. Plot of [$\rho$($t$) - $\rho$(0)]/$\rho$(0) with $t$ is illustrated in fig. 6(b) by the open circles where $\rho$(0) defines the resistivity at $t$ = 10 s. The relaxation of magnetization is involved with the activation against distribution of anisotropy barriers for the glassy magnetic compound where time dependence of magnetization typically follows the stretched exponential with a critical exponent ($\beta$) for 0 $<$ $\beta$ $<$ 1. We fit the time dependence using a stretched exponential function, [$\rho$($t$) - $\rho$(0)]/$\rho$(0) = $A + B\exp(t/\tau)^{\beta}$ where $\tau$ is the relaxation time. The satisfactory fit is shown by the continuous curve in fig. 6(b) with $\tau$ = 2435 s and $\beta$ = 0.94, suggesting the glassy magnetic behavior in the transport property. We observe a considerable increase of [$\rho$($t$) - $\rho$(0)]/$\rho$(0) up to $\sim$ 11 \% at $t$ = 5.6$\times 10^3$ s. On the other hand, a very small decrerase of [$\rho$($t$) - $\rho$(0)]/$\rho$(0) around $\sim$ 1 \% is observed below 100 s when the sample was cooled down to 5 K from 250 K in ZFC mode and then $\rho$($t$) was recorded with time in 50 kOe field. The results clearly demonstrate that strong time dependence of $\rho$ at 5 K in FC mode is involved with the EB effect. 

A cartoon of the phase separation within a grain is proposed in the left  inset of fig. 6(b) where short range FM regions or clusters are embedded in a SG matrix. When the sample is cooled in FC mode, a new layer consisting of frozen FM (FFM) spins arises  at the FM/SG interface due to the pinning of the FM spins by the rigid SG spins. The layers of FFM spins by replacing the surface region of the FM clusters at the vicinity of FM/SG interface are depicted in the right inset of fig. 6(b). The appearance of new layer causes the marked deviation between $\rho_{\rm FC}$ and  $\rho_{\rm ZFC}$ where $\rho_{\rm FC}$ is larger in magnitude than $\rho_{\rm ZFC}$ below 40 K. Moreover, the appearance of new layer leads to the strong time dependence in $\rho$ satisfying the stretched exponential which is typically observed in the time evolution of magnetization characterizing the glassy magnetic behavior. The results indicate that frozen FM spins pinned by the SG spins at the FM/SG  interface reveal the glassy magnetic behavior in the transport property, although it was originally FM spins before field cooling. 

Since the size and the density of FM clusters within the grain is close to the percolation threshold, tunneling among the FM metallic clusters takes place through the barrier involved with the anisotropy of SG spins. The tunneling barrier is further modified by the appearance of new layers comprising of FFM spins having unidirectional anisotropy which gives rise to the unidirectional shift in the MR-$H$ curve. Polarization direction of FFM spins are strongly influenced by the direction of cooling field leading to the spin valve mechanism in MR. We noticed that the decreasing MR-$H$ curve (from 50 kOe to -50 kOe) is modified considerably retaining nearly unchanged increasing MR-$H$ curve (from -50 kOe to 50 kOe) where the peak in the decreasing curve is shifted along negative $H$-axis [see figs. 4(c) and (d)] for positive cooling field. On the other hand, the increasing MR-$H$ curve is modified for the negative cooling field giving rise to the spin valve mechanism. We further note the marked difference between exchange bias field estimated from the different measurement techniques where anisotropies manifested by the coercivities are considerably different for the magnetization and resistivity measurements with $H_{\rm C}$ = 5.8 kOe and $H_{\rm C}^\prime$ = 8.8 kOe, respectively for the bulk. The simple intuitive model  \cite{meik1} as well as sophisticated theories \cite{wein,hu} seem to agree that the exchange bias effect should be larger for larger anisotropy of the AFM substance in a FM/AFM heterostructure which has been elaborately described by Nogu\'{e}s {\it el al}.  \cite{nogues2}. Although the experimental investigations \cite{taka,dij,lin} dealing with the role of the anisotropy seem to agree with the theories, any quantitative conclusions from them could not be established because of the difficulties for extracting the anisotropy of the AFM component from the heterostructure. In the present observation the larger EB  field is found to be  associated with the larger coercivity obtained from the MR-$H$ curve which is in accordance with the proposed theories \cite{meik1,wein,hu} as well as the experimental results \cite{taka,dij,lin}. Until now, different aspects of EB effect  has been extensively investigated through the magnetization studies which is rather less focused in the MR measurement. Thus a clear understanding of EB effect correlated with the coercivity of the highly anisotropic substance is still lacking at the microscopic level especially, for the MR measurements.

\section{Conclusions}

In conclusion, the compound exhibits large magnetoresistance which is involved with the intragranular spin polarized tunneling mechanism where tunneling barrier is further modified by the appearance of frozen ferromagnetic spins due to the field-cooling. Frozen ferromagnetic spins give rise to the glassy transport behavior and more importantly the exchange bias effect. Exchange bias field obtained from the magnetoresistance is much larger than the value of exchange bias field measured from magnetic hysteresis loop. Semiconducting transport properties associated with the spin polarized tunneling mechanism and considerable exchange bias  exhibiting spin valve like feature in bulk La$_{0.88}$Sr$_{0.12}$CoO$_3$ is attributed to the intrinsic nanostructure in the grain interior which creates a tremendous impact in searching new materials for the spintronics applications.  

\noindent
{\bf Acknowledgment}
S.G. wishes to thank DST, India for the financial support. M.P. thanks CSIR, India for the fellowship. 











\end{document}